\documentstyle[aps,prd]{revtex}
\tighten 

\begin{document}
\draft
\twocolumn

\wideabs{
\title{Chiral Superconducting Strings and Nambu-Goto Strings in 
Arbitrary Dimensions}

\author{Xavier Siemens and Ken D. Olum}

\address{Institute of Cosmology\\
Department of Physics and Astronomy\\
Tufts University\\
Medford MA 02155, USA}

\date{\today}

\maketitle

\begin{abstract}
We present general solutions to the equations of motion for a superconducting
relativistic chiral string that satisfy the unit magnitude constraint 
in terms of products of rotations. From this result we show how to construct 
a general family of odd
harmonic superconducting chiral loops. We further generalise the
product of rotations to an arbitrary number of dimensions.
\end{abstract}


}

\section{Preliminaries}

Particle physics models where symmetry breaking is involved predict, in many cases,
the existence of topological defects, which are formed when the topology of the
vacuum manifold of the low energy theory is non-trivial \cite{1}. Cosmic strings, in
particular, are line-like objects that are formed when the the vacuum manifold 
contains unshrinkable loops. For a review see \cite{2}.

In \cite{3} it was shown that cosmic strings can be superconducting. In the
case when the charge carriers on the string are not coupled to a gauge field 
the action for the string and the current can be taken to be
\begin{equation}
\label{action}
S=\int d^{2}\xi \sqrt{-\gamma }(-\mu +\frac{1}{2}\gamma ^{ab}\phi _{,a}\phi _{,b})
\end{equation}
where $\mu$ is the mass per unit length of the string, $\gamma^{ab}$ is the induced metric 
on the string worldsheet and $\phi$ is the field of the charge carriers living on the string.
These strings were shown in \cite{4} and \cite{5} to have solutions in the
case when \( \gamma ^{ab}\phi _{,a}\phi _{,b}=0 \) of the form
\begin{equation}
\label{stringmotion}
{\bf x}=\frac{1}{2}[{\bf a}(u)+{\bf b}(v)]
\end{equation}
for the string position and 
\begin{equation}
\label{current}
\phi =\frac{1}{2}f(v)
\end{equation}
for the field living on the string with the constraints 
\begin{equation}
\label{constrfora}
{\bf a}'^{2}=1
\end{equation}
and
\begin{equation}
\label{constrforbandf}
{\bf b}'^{2}+f'^{2}=1
\end{equation}
where \( u=\sigma -\tau  \) and \( v=\sigma +\tau  \) and \( \sigma  \)
and \( \tau  \) are space-like and time-like parameters respectively that parametrise
the string world-sheet. These strings are called chiral because the current
only moves in one direction on the string.

Comparing this to the usual Nambu-Goto case one can see that \( f(v) \) acts like
a fourth component of the three-vector \( {\bf b}(v) \), making chiral superconducting
strings behave like Nambu-Goto ones with chiral excitations in an extra fifth
dimension. Indeed, this property was used in \cite{6} in an investigation
of the properties of superconducting cosmic string cusps. 

The right- and left-moving excitations, \( {\bf a}' \) and \( {\bf b}' \),
on a regular Nambu-Goto string in Minkowski space-time are arbitrary functions
that satisfy the unit magnitude constraint, 
$\left| {\bf a}'\right|= \left|{\bf b}'\right|=1$. Expressions for these functions
are often given as Fourier sums and the unit magnitude constraint generally
gives a non-linear set of equations involving the vector coefficients of the
Fourier expansion. As a result, parametrising strings beyond the first few harmonics
proves to be a difficult task. Fortunately, in that case, there exists a method
to generate strings involving products of rotation matrices \cite{7} that act
on a starting unit vector so that the unit magnitude constraint is satisfied
trivially. 

In a recent study of the properties of chiral cosmic strings \cite{8} it was
assumed that the current is constant. The work in \cite{9} assumed that the
current takes a very simple non-constant form. As was pointed out in the latter work
one could expect to have loops with varying currents if the loops are formed by 
intersections
involving different strings or if different segments of the loop or string were
at some point in causally disconnected regions. For long strings this is always
the case and we therefore expect varying chiral currents to be generic.

The purpose of this work is to generalise the work in \cite{7} for generating
three dimensional unit vectors to four dimensional ones that include the current
as a fourth component of \( {\bf b}(\sigma +\tau ) \). We start by casting
the method somewhat differently and generalise it to four dimensions. From this
result we construct a family of chiral superconducting odd-harmonic loops. 
We further generalise the product of rotations 
to an arbitrary number of dimensions. 

In Section 2 we show how
to construct an arbitrary \( N \) harmonic unit vector in four and three dimensions.
In Section 3 we use these results to construct arbitrary chiral current
carrying superconducting odd harmonic loops.
In Section 4 we generalise the arguments in Section 2 
to an arbitrary number of dimensions and we conclude in Section 5. 

\section{Solution to the Unit Magnitude Constraint in Terms of Products of Rotations}

We can think of the Euclidean \( 4 \)-vector 
\begin{equation}
\label{4dbvector}
{\bf \tilde{b}}'=\left( \begin{array}{c}
b_{w}'\\
b_{x}'\\
b_{y}'\\
b_{z}'
\end{array}\right) 
\end{equation}
as having unit magnitude according to (\ref{constrforbandf}) with \( b_{w}=f \).
Consider the vector \( {\bf \tilde{b}}_N ' \), that can be constructed from a
finite sum of Fourier components,
\begin{equation}
\label{fourier}
{\bf \tilde{b}}_{N}'(v)={\bf Z}+\sum _{n=1}^{N}\{{\bf A}_{n}\cos nv+{\bf B}_{n}\sin nv\}.
\end{equation}
The Fourier coefficients satisfy the set
of \( 4N+1 \) non-linear relations derived in \cite{7} 
\begin{equation}
\label{cons1}
\sum ^{N}_{n=m-N}({\bf \alpha }_{n}\cdot {\bf \alpha }_{m-n}
-{\bf \beta }_{n}\cdot {\bf \beta }_{m-n})=4\delta _{m0}
\end{equation}
with $m=0,1,...,2N$,
\begin{equation}
\label{cons2}
\sum ^{N}_{n=m-N}({\bf \alpha }_{n}\cdot {\bf \beta }_{m-n}
-{\bf \beta }_{n}\cdot {\bf \alpha }_{m-n})=0
\end{equation}
with $m=1,...,2N$,
\begin{equation}
\label{alphabetaABn}
{\bf \alpha }_{n}={\bf \alpha }_{-n}={\bf A}_{n},\: 
{\bf \beta }_{n}=-{\bf \beta }_{-n}={\bf B}_{n},\: \: \: n\neq 0,
\end{equation}
and
\begin{equation}
\label{alphabetaAB0}
{\bf \alpha }_{0}=2{\bf Z},\: {\bf \beta }_{0}=0.
\end{equation}
These equations can be obtained from the constraint equation (\ref{constrforbandf}).
The total number of degrees of freedom in the coefficients in (\ref{fourier})
is \( 8N+4 \) so the remaining number of degrees of freedom after satisfying
the constraint is \( 4N+3 \). Below we show how to construct (\ref{fourier})
from a product of rotation matrices by generalising a modified version of the
three dimensional method presented in \cite{7} to four dimensions.

The constraint equations (\ref{cons1}) and (\ref{cons2}) for \( m=2N \) and
\( m=2N-1 \) are 
\begin{equation}
\label{meq2Ncons}
{\bf A}_{N}^{2}={\bf B}^{2}_{N},\: \: {\bf A}_{N}\cdot {\bf B}_{N}=0
\end{equation}
and
\begin{eqnarray}
\label{meq2Nm1cons}
\nonumber
{\bf A}_{N}\cdot {\bf A}_{N-1}-{\bf B}_{N}\cdot {\bf B}_{N-1}&=&0,
\\ 
{\bf A}_{N}\cdot {\bf B}_{N-1}+{\bf B}_{N}\cdot {\bf A}_{N-1}&=&0
\end{eqnarray}
respectively. It follows from (\ref{meq2Ncons}) that the highest harmonic is
a circle that lives on some arbitrary plane. Clearly, we can introduce coordinates
such that \( {\bf A}_{N} \) and \( {\bf B}_{N} \) lie on the \( w \) and
\( x \) axes 
\begin{equation}
\label{ANBN}
{\bf A}_{N}=a\hat{w},\: \: {\bf B}_{N}=a\hat{x}
\end{equation}
making the highest harmonic a circle of radius \( a \) on the \( w \)-\( x \)
plane. This puts the vector \( {\bf \tilde{b}}_{N}' \) into the so-called standard
form \cite{7}. Let $R_{wx}(\theta)$ be a matrix that rotates the $w$-$x$ plane by an angle $\theta$.
Acting on \( {\bf \tilde{b}}_{N}' \) with, \( R_{wx}(-v) \)
in these coordinates lowers the highest harmonic term of (\ref{fourier}), 
\begin{equation}
\label{RwxuonAB}
R_{wx}(-v)\left( \begin{array}{c}
a\cos Nv\\
a\sin Nv\\
0\\
0
\end{array}\right) =\left( \begin{array}{c}
a\cos (N-1)v\\
a\sin (N-1)v\\
0\\
0
\end{array}\right) .
\end{equation}
This is not sufficient to verify that the overall harmonic content has been
lowered because the \( N-1 \) terms of (\ref{fourier}) could still give us
an \( N \) harmonic through trigonometric identities. We now show, however,
that this is not the case. 

The constraint equations for \( m=2N-1 \) (\ref{meq2Nm1cons}) give us using
(\ref{ANBN}) the conditions on the coefficients,
\begin{equation}
\label{Anm1Bnm1}
({\bf A}_{N-1})_{w}=({\bf B}_{N-1})_{x},\: \: ({\bf A}_{N-1})_{x}=-({\bf B}_{N-1})_{w}.
\end{equation}
It is not too difficult to see that using the conditions (\ref{Anm1Bnm1})
when acting with \( R_{wx}(-v) \) on the \( N-1 \) terms of (\ref{fourier})
does not lead to the creation of \( N \) harmonic terms,
\begin{eqnarray}
\label{longasseqn}
&R_{wx}&(-v) \left[ 
{\bf A}_{N-1} \cos (N-1)v
+{\bf B}_{N-1}\sin (N-1)v \right]
\\
&=&\left( \begin{array}{c}
({\bf A}_{N-1})_{w}\cos (N-2)v-({\bf A}_{N-1})_{x}\sin (N-2)v\\
({\bf A}_{N-1})_{x}\cos (N-2)v+({\bf A}_{N-1})_{w}\sin (N-2)v\\
({\bf A}_{N-1})_{y}\cos (N-1)v+({\bf B}_{N-1})_{y}\sin (N-1)v\\
({\bf A}_{N-1})_{z}\cos (N-1)v+({\bf B}_{N-1})_{z}\sin (N-1)v
\end{array}\right) .
\nonumber
\end{eqnarray}
Clearly, the \( N-2 \) terms will not yield an \( N \) harmonic term when
acted on by \( R_{wx}(-v) \). 

We now have a string in the form of (\ref{fourier}) with \( N\rightarrow N-1 \).
Its highest harmonic is therefore also a circle of some other radius living
on some other arbitrary plane. Armed with this knowledge we can see that the
unit magnitude constraints are such that in general we can write
\begin{equation}
\label{bnm1plane}
{\bf \tilde{b}}_{N}'=R_{P_{N}}(v){\bf \tilde{b}}_{N-1}'
\end{equation}
where \( R_{P_{N}}(v) \) is a rotation by an angle \( v \) on the plane \( P_{N} \)
where the highest harmonic lives. By induction it must be that
\begin{equation}
\label{OurProdofRots}
{\bf \tilde{b}}_{N}'=\prod ^{1}_{i=N}R_{P_{i}}(v){\bf \tilde{b}}_{0}'
\end{equation}
where the \( R_{P_{i}}(v) \) are rotations by an angle \( v \) on arbitrary
planes \( P_{i} \) and \( {\bf \tilde{b}}_{0}' \) is an arbitrary constant
unit vector in \( 4 \) dimensions. 

To specify a plane in \( 4 \) dimensions one needs to specify a direction on
the plane (\( 3 \) angles), a linearly independent direction (\( 2 \)
angles), but now the plane is overspecified by internal rotations (\( -1 \)
angle), giving a total of \( 4 \) degrees of freedom for the matrices \( R_{P_{i}}(v) \).
For an \( N \) harmonic vector, one therefore has \( 4N \) parameters in the
rotators and \( 3 \) parameters in the constant unit vector \( {\bf \tilde{b}}_{0}' \)
giving a total of \( 4N+3 \) parameters which checks perfectly with \( 8N+4 \)
vector coefficients in (\ref{fourier}) minus the \( 4N+1 \) constraints (\ref{cons1})
and (\ref{cons2}). 

The form of the \( R_{P_{i}}(v) \) matrices is quite simple.
Generally, if we want to transform, say, a rotation by an angle \( v \) on
the $w$-$x$ plane to one on an arbitrary plane we will need to perform a
transformation of the type
\begin{equation}
\label{gentransf}
R_{P_{i}}(v)=E_{i}R_{wx}(v)E_{i}^{T}.
\end{equation}
For the purpose of finding the form of the rotators \( E_{i} \) it is easiest
to envision the inverse process to the one we are seeking, namely the rotation
of an arbitrary oriented plane to lie on the \( w \)-\( x \) plane. Let's consider
first the projection of the four dimensional arbitrary plane onto the \( (x,y,z) \)
subspace. We can perform a rotation by some angle \( \alpha  \) about the \( z \)
axis (\( R_{xy}(\alpha ) \)) until the vector perpendicular to the projected
plane lies on the \( y \)-\( z \) plane and perform a further rotation by
an angle \( \beta  \) about the \( x \) axis (\( R_{yz}(\beta ) \)) until
that vector lies on the \( z \) axis. At this stage the projected plane lies
wholly on the $x$-$y$ plane. The ranges of both \( \alpha  \) and \( \beta  \)
from \( 0 \) to \( \pi  \) are sufficient to perform these transformations.
After performing these two rotations, our original four dimensional plane lies
entirely in the \( (w,x,y) \) subspace and we can repeat an analogous process
to the one above to rotate it into the \( w \)-\( x \) plane. A rotation by
an angle \( \gamma  \) about the \( y \) axis (\( R_{wx}(\gamma ) \)) puts
the vector perpendicular to the plane on the \( x \)-\( y \) plane and a rotation
by an angle \( \delta  \) about \( w \) (\( R_{xy}(\delta ) \)) makes that
vector parallel with the \( y \) axis. For the first of these rotations a range
of \( \gamma  \) from \( 0 \) to \( \pi  \) is sufficient, for the second
rotation, however, matters are slightly different. If the plane we were trying
to rotate was featureless it would be enough for the range of \( \delta  \)
to be from \( 0 \) to \( \pi  \). In fact this is not the case. The plane
contains a circle in \( v \) which can be oriented clockwise or anti-clockwise
on the \( w \)-\( x \) plane and therefore the final rotation on the \( x \)-\( y \)
plane in general requires an angle that ranges from \( 0 \) to \( 2\pi  \).

Keeping in mind these considerations we can quite generally write
\begin{equation}
\label{rotator}
E_{i}=R_{xy}(\alpha _{i})R_{yz}(\beta _{i})R_{wx}(\gamma _{i})R_{xy}(\delta _{i})
\end{equation}
where \( \alpha _{i},\, \beta _{i},\, \gamma _{i} \) range from \( 0 \) to
\( \pi  \) and \( \delta _{i} \) ranges from \( 0 \) to \( 2\pi  \). Then
\begin{equation}
\label{OPOR1}
{\bf \tilde{b}}_{N}'(v)=\prod ^{1}_{i=N}E_{i}R_{wx}(v)E_{i}^{T}{\bf \tilde{b}}_{0}'.
\end{equation}

In order to construct the entire chiral string, we also need to find
the form of \( {\bf a}(u) \) in (\ref{stringmotion}). The constraints (\ref{constrfora})
can be satisfied using a product of rotations that can be found from analogous
arguments to the ones in the preceding section. For \( M \) harmonics this
yields
\begin{equation}
\label{3DProdofrots}
{\bf a}_{M}'(u)=\prod ^{1}_{i=M}D_{i}R_{xy}(u)D_{i}^{T}{\bf a}_{0}'
\end{equation}
with the rotator 
\begin{equation}
\label{3Drotator}
D_{i}=R_{xy}(\phi _{i})R_{yz}(\theta _{i})
\end{equation}
where the angles \( \theta _{i} \) range from \( 0 \) to \( \pi  \), the
angles \( \phi _{i} \) from \( 0 \) to \( 2\pi  \) and \( R \) are
the three dimensional rotation matrices.

\section{An Application: A Family of 
Odd Harmonic Superconducting Chiral Loops}

\subsection{Overall Orientation Freedom}

Both expressions for the oppositely moving excitations on the string (\ref{OPOR1})
and (\ref{3DProdofrots}) include overall orientation freedom. In some applications, 
for instance self-intersection or gravitational radiation analyses, only the shape of a loop, 
and not its orientation, is important. 
In this case the inclusion of overall orientation freedom of
the right-moving and left-moving excitations separately is unnecessary: All
that matters is the relative orientation between \( {\bf a}' \) and the spatial
part of \( {\bf \tilde{b}}' \). 

Overall orientation freedom of the loop is
set by an Euler matrix \( Q \) that acts only on  \( {\bf a}' \) and the spatial components 
of \( {\bf \tilde{b}}' \) and contains three angles. We can use this freedom 
to standardise the vectors in some way. We will choose the plane of the last rotation 
in \( {\bf a}' \) and \( {\bf \tilde{b}}' \) to contain some coordinate axis 
(this will enable us in the following sub-section to use the arbitrariness in the 
origin of $u$ and $v$ to eliminate more parameters).

The four-dimensional 
circle that constitutes the
highest harmonic of \( {\bf \tilde{b}}' \) projected onto three-dimensional space
typically looks like an ellipse. 
We can perform ordinary three-dimensional rotations on it to put 
it in some convenient form. In particular, we can rotate the ellipse 
on the $x$-$y$ plane until its major axis, say, lies on the $y$-$z$ 
plane and further rotate on the $y$-$z$ plane until it lies entirely 
on the $z$ axis. We still have one more rotation left in $Q$ which 
we choose to be a rotation on the $x$-$y$ plane. This leaves the 
major axis of the ellipse on the $z$-axis but orients the circle of 
the highest harmonic in \( {\bf a}' \) such that it contains the $x$-axis.

We now construct a four-dimensional planar rotator 
that contains the $z$-axis but is otherwise arbitrary. 
Starting with $R_{wz}(v)$, 
we see that we can perform an arbitrary planar rotation on it in the $(w,x,y)$ subspace
that preserves the $z$-axis. This requires only two rotations. Therefore, to 
rotate about a plane that contains the $z$-axis but is otherwise arbitrary we can 
use 
\begin{equation}
\label{rot0b}
R_{xy}(\gamma)R_{wx}(\delta)R_{wz}(v)R_{wx}(-\delta)R_{xy}(-\gamma)
\end{equation}
where $\delta$ ranges from $0$ to $\pi$ and $\gamma$ from $0$ to $2\pi$.
This procedure eliminates two of the parameters in the last planar 
rotation of  \( {\bf \tilde{b}}' \).

In three dimensions to rotate by an angle $u$ on a plane that contains 
the $x$-axis,
but is otherwise arbitrary, we may use
\begin{equation}
\label{rot0a}
R_{yz}(\theta)R_{xy}(u)R_{yz}(-\theta)
\end{equation}
where $\theta$ ranges from $0$ to $2\pi$.
This procedure eliminates one of the parameters in the last planar 
rotation of  \( {\bf a}' \).

This leaves  \( {\bf a}' \) and 
\( {\bf \tilde{b}}' \) in the form  
\begin{equation}
\label{OPOR2b}
{\bf \tilde{b}}_{N}'(v)={\tilde R}^{b}_{P_{N}}(v)
\prod ^{1}_{i=N-1}E_{i}R_{wx}(v)E_{i}^{T}{\bf \tilde{b}}_{0}'.
\end{equation}
with
\begin{equation}
\label{Rnbtilde}
{\tilde R}^{b}_{P_{N}}(v)=R_{wx}(\gamma)R_{xy}(\delta)R_{wz}(v)R_{xy}(-\delta)R_{wx}(-\gamma)
\end{equation}
and
\begin{equation}
\label{OPOR2a}
{\bf a}_{M}'(u)={\tilde R}^{a}_{P_{M}}(u) 
\prod ^{1}_{i=M-1}D_{i}R_{xy}(u)D_{i}^{T}{\bf a}_{0}'
\end{equation}
with
\begin{equation}
\label{Rnatilde}
{\tilde R}^{a}_{P_{M}}(u)=R_{yz}(\theta)R_{xy}(u)R_{yz}(-\theta).
\end{equation}

\subsection{The Origin of $u$ and $v$}

The conditions on the coefficients of the highest harmonics (\ref{meq2Ncons}) 
only specify the planar rotation up to a phase so that generally we can take 
${\tilde R}^{b}_{N}(v+\beta)$ and ${\tilde R}^{a}_{M}(u+\alpha)$ 
in (\ref{OPOR2b}) and (\ref{OPOR2a}).

We consider the action of this extra planar rotation matrix 
on \( {\bf \tilde{b}}_{N}'(v) \). It can be verified to be
\begin{equation}
\label{Rtildebprime}
{\bf \tilde{b}}_{N}'(v)={\tilde R}^{b}_{P_{N}}(v)
\prod ^{1}_{i=N-1} R_{P'_{i}}(v){\tilde R}^{b}_{P_{N}}(\beta){\bf \tilde{b}}_{0}'.
\end{equation}
where 
\begin{equation}
\label{Rpp}
R_{P'_{i}}(v)={\tilde R}^{b}_{P_{N}}(\beta)R_{P_{i}}(v){\tilde R}^{b}_{P_{N}}(-\beta) 
\end{equation} 
with analogous expressions for
\( {\bf a}_{M}'(u) \). It is important to note that the effect of 
replacing $ R_{P_{i}}(v)$ with $R_{P'_{i}}(v)$ 
is to make the same transformation on each of the planes of rotation, in other words, to 
rotate on some other set of planes. Since we can express any rotation on a 
plane using (\ref{rotator}) however, the effect of the matrices  
${\tilde R}^{b}_{P_{N}}(\beta)$ and ${\tilde R}^{a}_{P_{M}}(\alpha)$ on the 
rotators can be ignored.

Since the planar rotations ${\tilde R}^{b}_{P_{N}}$ 
and  ${\tilde R}^{a}_{P_{M}}$ both contain one of the coordinate axes we
can choose $\beta$ and $\alpha$ so that 
the $z$ component of ${\bf \tilde{b}}_{0}'$ and 
the $x$ component ${\bf {a}}_{0}'$ vanish. This leaves them in the form
\begin{equation}
\label{Rtildeb0prime}
{\tilde R}^{b}_{P_{N}}(\beta){\bf \tilde{b}}_{0}'
=\left( \begin{array}{c}
\cos \theta_b\\
\cos \phi_b \sin \theta_b\\
\sin \phi_b \sin \theta_b\\
0
\end{array}\right)=R_{xy}(\phi_b) R_{wx}(\theta_b) {\hat w}
\end{equation}
and
\begin{equation}
\label{Rtildea0prime}
{\tilde R}^{a}_{P_{M}}(\alpha){\bf {a}}_{0}'=\left( \begin{array}{c}
0\\
\cos \theta_a\\
\sin \theta_a\\
\end{array}\right)=R_{yz}(\theta_a) {\hat y}
\end{equation}
where $\phi_b$ and $\theta_a$ range from $0$ to $2\pi$ and $\theta_b$ 
ranges from $0$ to $\pi$.
This means we can write  \( {\bf a}' \) and \( {\bf \tilde{b}}' \) as
\begin{equation}
\label{OriginandOrientaprime}
{\bf a}_{M}'(u)={\tilde R}^{a}_{P_{M}}(u)
\prod ^{1}_{i=M-1} R_{P_{i}}(u)R_{yz}(\theta_a){\hat y}.
\end{equation}
and
\begin{equation}
\label{OriginandOrientbprime}
{\bf \tilde{b}}_{N}'(v)={\tilde R}^{b}_{P_{N}}(v)
\prod ^{1}_{i=N-1} R_{P_{i}}(v)R_{xy}(\phi_b) R_{wx}(\theta_b) {\hat w}.
\end{equation}

\subsection{The Center of Mass Constraint}

In order to construct string loops, apart from solving the unit magnitude constraint,
we need to satisfy the center of mass constraint, namely that the loop should be closed
and that we want to work
in the rest frame
of the loop. These constraints imply that the center of mass term must 
vanish, ${\bf Z}=0$ in (\ref{fourier}), for both ${\bf a}'$ and ${\bf {\tilde b}}'$. 
In general this is an intractable
problem because the center of mass term \( {\bf Z} \) is
a non-linear function of the angles in the rotation matrices. We can, however, solve 
this problem in an analogous way to that introduced in \cite{10}. The guiding principle
behind such a construction is to set the starting center of mass terms to zero,
and to apply further rotation matrices in such a way as to ensure that trigonometric
identities never lead to the production of a zero harmonic. In practice this
was done by using odd harmonics only and choosing some parameters so that the
starting center of mass term is zero. Here we will proceed along similar lines.

Generally, if the starting unit vector lies somewhere on the plane of the first 
rotation, \( i=1 \) in (\ref{OriginandOrientaprime}) 
and (\ref{OriginandOrientbprime}), no center of mass term will
be generated. Since our starting unit vectors are given by (\ref{Rtildeb0prime})
and (\ref{Rtildea0prime}) we need to choose the first planes of rotation appropriately.

Earlier we established that if we want a rotation by an angle $u$ on a plane that 
contains the $x$-axis we can use (\ref{rot0a}). If we want an arbitrary rotation 
that contains the $y$-axis instead we may use
\begin{equation}
\label{rot1a}
R_{xz}(\phi)R_{xy}(u)R_{xz}(-\phi).
\end{equation}
If we now decide we want a rotation by $u$ 
on a plane that contains a vector lying somewhere on the $y$-$z$ plane, it is 
sufficient rotate this last matrix (\ref{rot1a}) on the $y$-$z$ plane by whatever 
angle the vector makes with the $y$-axis. This is precisely the 
situation in (\ref{Rtildea0prime}). We therefore
write the first rotation, $i=1$ in (\ref{OriginandOrientaprime}) as
\begin{equation}
\label{rot2a}
R_{P_{1}}(u)=R_{yz}(\theta_a)R_{xz}(\phi_1)
R_{xy}(u)R_{xz}(-\phi_1)R_{yz}(-\theta_a)
\end{equation}
where $\phi_1$ ranges from $0$ to $2\pi$.

To find the first rotator in (\ref{OriginandOrientaprime}) we proceed analogously. 
We want to find a rotation by an angle $v$ on a plane that contains a vector
lying somewhere in the \( (w,x,y) \) subspace, but is otherwise arbitrary,
because this is the situation 
of (\ref{Rtildeb0prime}). If we start with a rotator that contains the $w$-axis but
is otherwise arbitrary, taking say,
\begin{equation}
\label{rot1b}
R_{yz}(\gamma_1)R_{xy}(\delta_1)R_{wx}(v)R_{xy}(-\delta_1)R_{yz}(-\gamma_1),
\end{equation}
where $\delta_1$ ranges from $0$ to $\pi$ and $\gamma_1$ ranges from $0$ to $2\pi$,
it is not hard to see that the first rotator in (\ref{OriginandOrientaprime}) must be
\begin{eqnarray}
\label{rot2b}
R_{P_{1}}(v)&=&
R_{xy}(\theta_b)R_{wx}(\phi_b)
R_{yz}(\gamma_1)R_{xy}(\delta_1)R_{wx}(v)
\nonumber
\\
&\times&
R_{xy}(-\delta_1)R_{yz}(-\gamma_1)
R_{wx}(-\phi_b)R_{xy}(-\theta_b).
\end{eqnarray}

Further rotations should be by \( 2u \) and \( 2v \) to avoid the production
of center of mass terms through trigonometric identities. These considerations
yield
\begin{eqnarray}
\label{OddharmProdofrots}
{\bf \tilde{b}}_{2N-1}'(&v&)=
{\tilde R}^{b}_{P_{N}}(2v)
\prod ^{2}_{i=N-1}E_{i}
R_{wx}(2v)E_{i}^{T} 
\nonumber
\\
&\times& R_{xy}(\theta_b)R_{wx}(\phi_b)R_{xz}(\gamma_1)R_{xy}(\delta_1)R_{wx}(v)\hat{w}
\end{eqnarray}
and
\begin{eqnarray}
\label{OddharProfofrotd3D}
{\bf a}_{2M-1}'(u)=
{\tilde R}^{a}_{P_{M}}(&2u&)
\prod ^{2}_{i=M}D_{i}R_{xy}(2u)D_{i}^{T}
\nonumber
\\
&\times& 
R_{xy}(\theta_a)R_{yz}(\phi_1)R_{xy}(u)\hat{y}.
\end{eqnarray}

\section{Generalisation to Higher Dimensions}

Here we show the straightforward generalisation of our four dimensional
argument in Section 2 to a unit vector
living in arbitrary dimensions. The constraint equations (\ref{cons1}) and
(\ref{cons2}) as well as the argument leading to (\ref{OurProdofRots}) and
(\ref{gentransf}) are independent of the number of dimensions the vector lives
in. The only thing that changes with the number of dimensions is the parametrisation
of \( E \), the rotator that takes an arbitrary oriented plane to the plane
of the first two coordinates. In the following, we label our spatial coordinates 
by the numbers \( 1 \) through \( d \).

If we consider the projection of the plane onto the subspace given by the last
three coordinates \( (d-2,d-1,d) \) one can see that a rotation by \( \alpha _{d} \)
about the \( d \) axis (\( R_{d-2,d-1}(\alpha _{d}) \)) until the vector perpendicular
to the projected plane lies in the $d$-$d-1$ plane followed by a rotation by an angle
\( \beta _{d} \) about the \( d-2 \) axis (\( R_{d-1,d}(\beta _{d}) \)) is
sufficient to rotate the projected plane out of the \( d \) axis. To perform
these transformations it is sufficient for the angles to range from \( 0 \)
to \( \pi  \). We can repeat this procedure by moving up \( d-2 \) times in
the coordinates until the plane lies entirely in the $1$-$2$ plane as desired,
ensuring that the range of the angle in the very last rotation \( R_{2,3}(\beta _{3}) \)
is from \( 0 \) to \( 2\pi  \) to account for the fact that we are dealing
with an oriented plane. We can then write the rotator as
\begin{eqnarray}
\label{arbitRotor}
\hspace{0cm}
E=R_{d-2,d-1}(\alpha _{d})R_{d-1,d}(\beta _{d})
... R_{1,2}(\alpha _{3})R_{2,3}(\beta _{3}).
\end{eqnarray}
The number of parameters introduced by such a product is \( 2(d-2) \) per harmonic
plus \( (d-1) \) parameters to specify the initial unit vector giving a total
of \( 2N(d-2)+d-1 \) which checks with \( d(2N+1) \) degrees of freedom in
the vector coefficients of the Fourier series minus \( 4N+1 \) constraints.

In the case of Nambu-Goto strings in an arbitrary number of dimensions \( d \)
one would span both right and left moving excitations according to
\begin{equation}
\label{opor2}
{\bf a}_{N}'(u)=\prod ^{1}_{i=N}E_{i}R_{1,2}(u)E_{i}^{T}{\bf a}_{0}'
\end{equation}
and
\begin{equation}
\label{opor3}
{\bf b}_{N}'(v)=\prod ^{1}_{i=N}E_{i}R_{1,2}(v)E_{i}^{T}{\bf b}_{0}'
\end{equation}
with the \( E_{i} \) given by the a choice of (\ref{arbitRotor}) appropriate
to the desired number of dimensions. 

\section{Conclusions}

We have generalised the solution to the unit magnitude constraint presented
in \cite{7} from three to four dimensions, casting it somewhat differently,
in an effort to arrive at a general parametrisation of chiral superconducting
strings with a finite number of harmonics. We have further shown how to 
construct loop solutions that satisfy the center of mass constraint and exclude
overall orientation freedom. This result is useful because in studies of the 
properties of chiral loops, such as self-intersection and gravitational 
radiation properties, overall orientation of the loop is unimportant.

Studies of chiral cosmic string loops with constant currents \cite{8} and simple
varying currents \cite{9} have been performed. Generally, however, we expect
the current to be arbitrarily varying when loops are formed by intersections
involving different strings or if different segments of the loop or string were
at some point in causally disconnected regions. This is a fairly generic situation
and a study of the properties of more general chiral loops should account for
these variations.

Along the way, we have found that our modification of the method lends itself
readily to a generalisation to arbitrary dimensions. We use such a generalisation
to present solutions that could be useful in the investigation of classical
relativistic strings in higher dimensions as well as strings in \( 3+1 \) Minkowski
space with currents and charges induced by Kaluza-Klein compactification \cite{11} 
when the back-reaction from the gauge fields can be considered negligible.\\

{\bf Acknowledgments}\\
We would like to thank Jose Juan Blanco-Pillado, Allen Everett, 
Alex Vilenkin and Benjamin Wandelt for fruitful discussions.


\begin{thebibliography}{}
\bibitem{1}T.W.B. Kibble, J. Phys. A{\bf 9}, 1387 (1976).
\bibitem{2}A. Vilenkin and E.P.S Shellard, Cosmic strings and other 
Topological Defects.
(Cambridge University Press, 1994).
\bibitem{3}E. Witten, Nucl. Phys. B{\bf 249}, 557 (1985).
\bibitem{4}B. Carter and P. Peter, Phys. Lett. B{\bf 466}, 41 (1999).
\bibitem{5}J.J. Blanco-Pillado, K.D. Olum and A. Vilenkin, Phys. Rev. 
D{\bf 63}, 103513 (2001).
\bibitem{6}K.D. Olum, J.J. Blanco-Pillado and X. Siemens, Nucl. Phys. 
B{\bf 599}, 446 (2001).
\bibitem{7}R.W. Brown and D.B. DeLaney, Phys. Rev. Lett. {\bf 63}, 474 
(1989); R.W. Brown, M.E.
Convery and D.B DeLaney, J. Math. Phys. {\bf 32}, 1674 (1991).
\bibitem{8}A.C. Davis, T.W.B Kibble, M. Pickles and D. Steer, Phys. Rev. 
D{\bf 62}, 083516 (2000).
\bibitem{9}D.A. Steer, Phys. Rev. D{\bf 63},  083517 (2001).
\bibitem{10}X. Siemens and T.W.B Kibble, Nucl. Phys. B4{\bf 38}, 307 (1995).
\bibitem{11}N.K. Nielsen, Nucl. Phys. B{\bf 167}, 249 (1980).
\end{thebibliography}
\end{document}